\documentclass[journal]{IEEEtran}
\usepackage{graphicx}
\usepackage{cite}
\usepackage{caption}
\usepackage{subfig}
\usepackage{float}

\ifCLASSINFOpdf
\else
\fi
\usepackage{fixltx2e}
\usepackage{gensymb}
\usepackage{stfloats}
\usepackage{amsmath}

\begin{document}
%
\title{A review for Tone-mapping Operators on Wide Dynamic Range Image}
%
%
%

\author{Ziyi~Liu\\
University of Calgary\\

}

\maketitle

\begin{abstract}
The dynamic range of our normal life can exceeds 120 dB, however, the smart-phone cameras and the conventional digital cameras can only capture a dynamic range of 90 dB, which sometimes leads to loss of details for the recorded image. Now, some professional hardware applications and image fusion algorithms have been devised to take wide dynamic range (WDR), but unfortunately existing devices cannot display WDR image. Tone mapping (TM) thus becomes an essential step for exhibiting WDR image on our ordinary screens, which convert the WDR image into low dynamic range (LDR) image. More and more researchers are focusing on this topic, and give their efforts to design an excellent tone mapping operator (TMO), showing detailed images as the same as the perception that human eyes could receive. Therefore, it is important for us to know the history, development, and trend of TM before proposing a practicable TMO. In this paper, we present a comprehensive study of the most well-known TMOs, which divides TMOs into traditional and machine learning-based category.
\end{abstract}

\begin{IEEEkeywords}
Tone mapping, wide dynamic range image, image processing, machine learning.
\end{IEEEkeywords}

%
\IEEEpeerreviewmaketitle

\section{Introduction}
\begin{figure}[h]
\centering
\includegraphics[width=\linewidth]{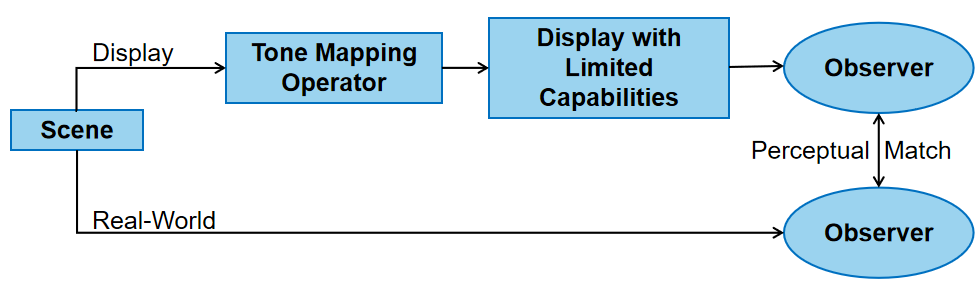}
\caption{Ideal tone mapping process}
\label{fig:1}
\end{figure}

The dynamic range of a scene is defined by the ratio of the light intensity of the brightest and the darkest spots. In the real-world, the dynamic range is extremely broad from 0 to 1, 000, 000. For the human visual system (HVS), we can perceive a range of around 24 EV when we view an outlook. Nevertheless, digital cameras can only capture a range of approximately 9 EV. If the dynamic range is more than 9 EV, there is no combination of aperture or shutter speed that will enable us to capture the entire dynamic range of the real scene. All you can do is optimize the exposure time for the highlight details or the shadow details, but you will inevitably miss one or the other. As a consequence, a wide dynamic range (WDR) image is proposed \cite{yadid1999wide}, \cite{spivak2009wide}. Unlike the low dynamic range (LDR) image, it allows you to record the darkest and brightest area of a scene at the same time that your camera wouldn’t be equal to capture in a single shot. Although WDR images (16/32-bit) have the ability to provide fascinating images, they are unable to be displayed via most regular devices (8-bit), and required to be transferred into LDR first. This process is called tone-mapping (TM), and is widely adopted in image processing and computer graphics nowadays. TM is a technique that is employed to convert a set of colors to another, approximating the WDR image information in instruments that have a limited dynamic range. An ideal tone mapping process is shown in Fig. \ref{fig:1}. It solves the problem of sharply reducing contrast from scene radiation to the displayable range, while retaining the image details and color appearances that are critical to appreciating the original scene content. Tone mapping originates from art where artists make full use of finite color to describe high contrast natural scenes. Afterward, it was launched for the application on television and photography. It leverages the fact that the HVS has a greater sensitivity to relative rather than absolute luminance levels \cite{barris2005vision}. The rest of the paper is organized as follows: The second section reviews 6 traditional TMOs: 3 global TMOs and 3 local TMOs. In the third section, we discusses 4 learning-based TMOs. We show the comparison and conclusion in section five and six separately.

\section{Traditional TMOs}
Global TMOs do mapping operations on all the pixels, thus they are very express compared to local TMOs. Most global TMOs cannot maintain image contrast. On the other hand, local TMOs apply variations on each pixel, taking properties of the neighboring pixels into account, but they are slower and might bring about artifacts in the output images. 

\subsection{Global TMOs}
Ward et al. \cite{ward1994contrast} designed a quite simple tone-mapping operator, utilizing a calculated scaling factor on the whole input WDR image to preserve contrast instead of absolute luminance. The method is based on subject studies of Blackwell \cite{blackwell1981analytical}, which established an association between adaptation luminance and minimum noticeable difference in brightness. From this study, Ward assumed that the result of adaptation can be seen as a shift in the absolute difference in brightness required for the viewer to notice the change, which means the visible luminance differences in the real-world can be mapped as the luminance differences on the display medium. After this operation, the contrast between the displayed image and the real image is fixed. This method is computationally efficient because it just applies a specific scaling factor to the input images and displays the resulting outputs, leading to the loss of visibility as a consequence of the clipping of the highest and the lowest pixel values. \\
Ferwerda et al. \cite{ferwerda1996model} developed a TM model derived from the psychophysical experiments they did for inferring changes in visual function that accompany light adaptation. The model simulates the variation of sensitivity, color appearance, visual acuity, and visibility as a function of the light level. Deploying this model could map the just noticeable differences (JNDs) in the real-world into JNDs on the medium, ignoring the unnecessary luminance values in input images. This model is very important due to the experiments, covering the thorough visual field via providing an immersive display system to such an extent the viewer's visual state could be determined on the entire display \cite{mcnamara2001visual}, but it does not fully capture the early stages of visual adaptation.
\\
A histogram adjustment TMO is proposed with Ward et al. \cite{larson1997visibility}, building on the earlier work \cite{ward1994contrast},\cite{ferwerda1996model}. First, they got the cumulative distribution of brightness histograms that is executed to identify clusters of light levels. With the aim of overcoming the dynamic range limitation, the author employed a linear function on an entire image, doing a histogram adjustment based on the gained distribution. This adjustment is established on the principles: luminance is not constant across a full image, even so it is consistent in a small region; human eyes are sensitive to color contrast instead of absolute brightness, only if the bright area is still brighter than the dim region in the processed resulting image, the absolute values are not important; eyes adapt rapidly to a small angle in the field of view (i.e., about 1\degree) from the fixation point. This method shows more overall contrast reproduction than the others, at the same time, it could guarantee that the contrast of the resulting image does not exceed the perception of the receptor. It also avoids the clipping problem of single adaptation methods. In spite of that, it still suffers the loss of information.

\subsection{Local TMOs}
Schlick et al. \cite{schlick1995quantization} came up with a quantization technique, using a so-called rational mapping function to transfer the WDR input into the displayable values. The quantization function has three versions, the first and the second one is just a simple logarithmic and exponentiation mapping respectively, which are hard to present a satisfying outcome. The third method has a rational mapping curve with asymmetry shape, treating the high and low pixels in a reciprocal way, thereby providing smoother results. They also introduced an approach for color treatment, the dynamic range is initially compressed in the luminance channel, and colors are reproduced in post-processing operations. All pixel values are got through a color ratio:
\begin{equation}
C_{out} = \frac{C_{in}}{L_{in}}L_{out}
\end{equation}

Where $C$ represents a color channel (red, green, or blue) of a colorful image, $L$ represents the luminance level image, and in/out denotes the input WDR or tone-mapped WDR image.
Even though this algorithm reduces the computation resources and parameter number, it is easily affected by halo artifacts. 
\\
Durand et al.’s \cite{durand2002fast} operator decomposed the input WDR image into two separate layers: a base layer and a detail layer, using a non-linear edge-preserving filter known as the bilateral filter \cite{tomasi1998bilateral}. The contrast of the base layer is decreased while preserving the details in the detail layer. The detail layer is further polished to elevate the small-scale details. To fast the decomposition process, they adopted two strategies: a piecewise-linear approximation in the intensity domain, and a sub-sampling in the spatial domain. At last, the tone-mapped WDR image can be obtained over combining processed two layers. It solves the halo artifacts that happened in \cite{schlick1995quantization}, but occasionally it over-enhances the local details of the image, losing realism.
\\
Jie et al. \cite{yang2018local} invented a TMO in a pixel-by-pixel manner, and they also designed tailored hardware to implement this TM algorithm. For the TMO, Jie first split the WDR input into mxn blocks, then assign a maximum and minimum operation on each block, obtaining two matrices (Mmax and Mmin). After a bilinear interpolation, the matrices are expanded into the same dimension as the input WDR image. The final equation is below:

\begin{multline} 
\label{eqn:2}
	d(i,j) = \frac{log(p(i,j))-log(M_{min}(i,j))}
		 {log(M_{max}(i,j))-log(M_{min}(i,j))}    \\
	\times (D_{max}-D_{min})+D_{min}
\end{multline}

Where the $p(i,j)$ represents the original pixel value in $(i,j)$ position of the input WDR image, $Dmax$ and $Dmin$ are the maximum and minimum display levels of the visualization tools. Using the logarithmic function and interpolation method could take the neighboring pixels into account, making full use of local information. The hardware implementation is made up of six modules pairing the modules of the TM approach. The parameter module collects the customer parameters such as width, height of image, while the pixel status module saves the location $(i, j)$ of one pixel and transfers the location into block div
module where the maximum and minimum of each image block are calculated. The calculated results are stored in the RegFile module, then interp module does a bilinear interpolation for the maximum and minimum. Eventually, eq. \ref{eqn:2} will be executed in the compute module. 
This method is able to get brighter images and is energy efficient. 
\section{Machine Learning-based TMOs}
\subsection{General Background}
\begin{figure}[h]
\centering
\includegraphics[scale = 0.40]{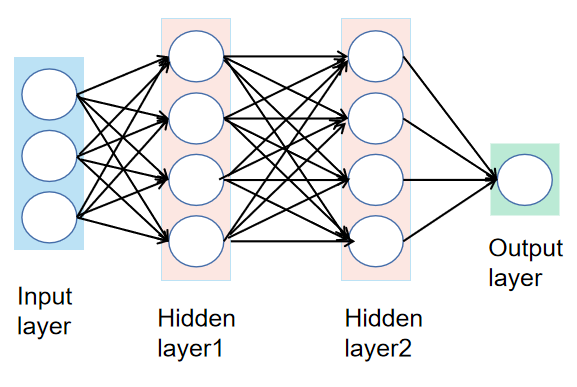}
\caption{Convolutional neural network architecture}
\label{fig:2}
\end{figure}

\begin{figure}[!h]
\centering
\includegraphics[scale = 0.30]{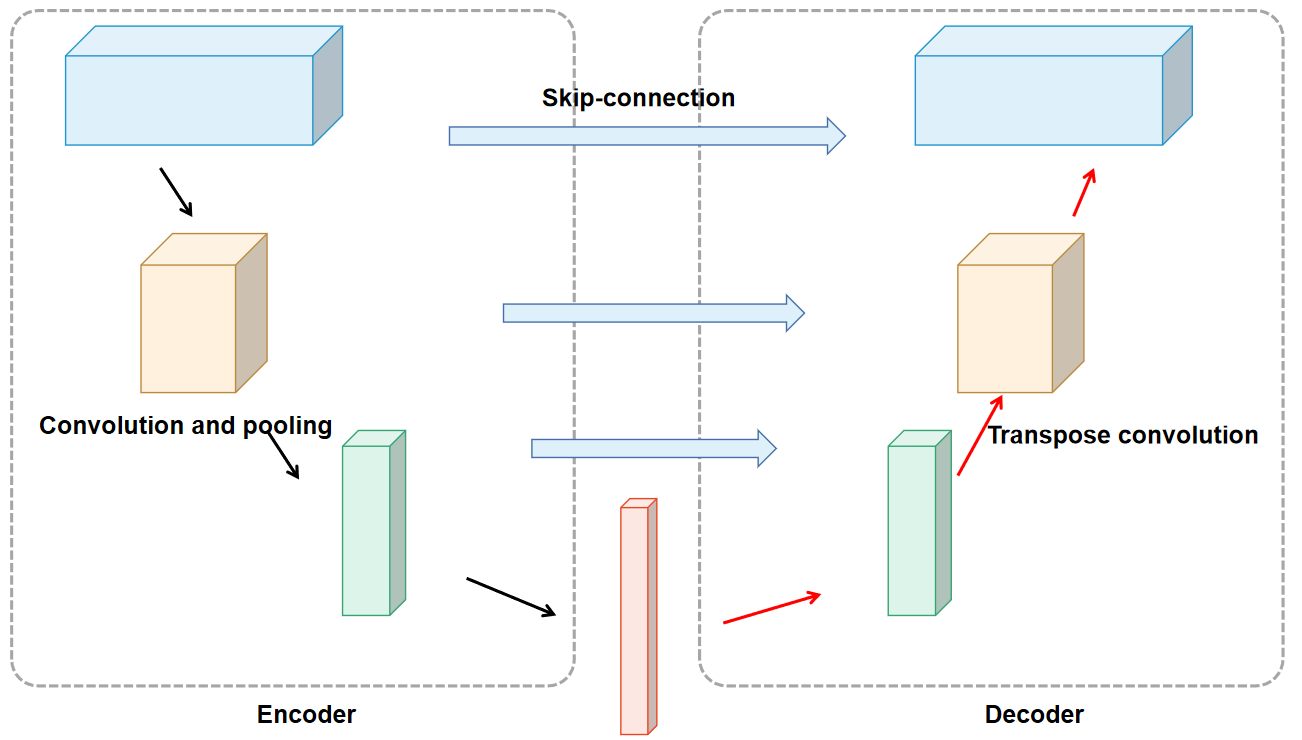}
\caption{U-Net architecture}
\label{fig:3}
\end{figure}

\textbf{Convolutional Neural Network.}
Convolutional Neural Networks (CNNs) are a sort of neural networks that do a mathematical operation termed convolution on at least one layer, instead of the common matrix multiplication. Usually, a CNN as shown in Fig. \ref{fig:2} includes an input layer, an output layer, as well as several specific hidden layers such as convolutional layers.\\
With the achievements of the CNNs on image classification \cite{krizhevsky2017imagenet}, medical image segmentation \cite{milletari2016v}, object detection \cite{ren2016faster}, natural language processing \cite{hu2014convolutional}, researchers begin to apply this knowledge to image-to-image translation tasks. Sharing weights reduces the complexity of CNNs, especially for an image with a multidimensional input vector, that can be directly inputted to the network, avoiding the complexity of data reconstruction in the process of feature extraction and classification. Nevertheless, CNN needs a specific loss function to understand how to minimize the differences between original images and their labeled images, sometimes this loss function potentially will not be reliable. For example, the L2 loss function would cause blurry in displayed images because it mixes many different types of attributes of input images, which can not make the image more realistic, but as far as the L2 loss function is concerned, this is the best image our CNN model could construct.\\
\textbf{Generative Adversarial Network.}
Generative Adversarial Network (GAN) is also a kind of neural networks. It is good at mimicing images. CNNs are good at extracting features of images, and the quality of resulting outputs of CNNs heavily relies on the choice of the loss function, taking a very long time to manually look for or design an appropriate loss function. Many solutions have been devised to fix this issue, among which GAN is the most prevalent one. It consists of two parts, one is a generator, another is the discriminator. As for the generator, it seems like a painter, keeping painting images, and its goal is to draw a masterpiece like Mona Lisa's Smile. In the meantime, the discriminator is like a connoisseur of art who has the ability to distinguish whether a painting is a real image or not, in other words, if a fake image painted in the generator is given to the discriminator, the discriminator will punish the generator until the generator could create a painting very similar to the reality that the discriminator cannot even differentiate. In order to achieve the same function as what GAN did, we conventionally have to design a complex mathematical loss function, but now we can leverage GAN to train its discriminator to satisfy the training data automatically. Nevertheless, this method cannot control the pattern of the data to be generated in GAN. Conditional GAN (CGAN) changes this condition by adding the tag Y to the generator as an additional parameter and wants to generate the corresponding image. CGAN could control the generated data conditionally dependent on the tag. This constrains the output space and minimizes the differences between the input and output images. The conditional information may possibly give a significant head start to GAN for what to look for, therefore, the productions of CGAN are perceived to be better.\\
\textbf{U-Net.}
U-Net \cite{bermudez2018domain} is a prevalent neural network model for biomedical image segmentation \cite{ronneberger2015u}, image translation problem \cite{wu2018deep}, due to the fact that U-Net advances processing and learning pixel-level information. It consists of an encoder and decoder with some skip-connections shown in \ref{fig:3}, which are applied for extracting features and reconstructing image accordingly. Encoder-decoder model was first proposed to be applied to medical image segmentation, and then it attracted people’s attention. Now it is universally used in image conversion, language translation, and other fields. Encoder is implemented to extract features from the input data (such as images, audio signal), converting the information from low-dimensional to high-dimensional, and obtaining the core information of the data. Decoder transforms the information features  into data in the target field, which is to map the high-dimensional core information to the low-dimensional target space. The encoder-decoder model structure generally incorporates a convolution layer, an upsampling layer, a deconvolution layer, and a pooling layer.\\
The high-dimensional image features extracted with the encoder are less sensitive to the content information in the image, but retain a large amount of semantic information of the image, which is beneficial to tasks such as image segmentation and object detection. The skip connection in this model, as the name implies, skips certain layers in the neural network, and provides the outputs of one layer as inputs to the next layer and other layers. Some information captured in the initial layer needs to be reconstructed when using the fully connected layer for up-sampling. If we do not execute skip connection, the information will be lost (or it should be said that it will become too abstract for further use). Therefore, the information we have in the main layer can be explicitly provided to the following layers using a skip connection structure.\\

\textbf{Supervised learning vs Unsupervised learning.}
In supervised learning, the neural network is trained on a labeled database, learning how to map input data into the ground truth (GT) in the labeled database, whereas unsupervised learning dosed not provide a labeled database, the machine learning model needs to figure out the features, patterns, and relationship among these unlabeled data on their own. 

\subsection{TMOs}
\begin{figure}[h]
\centering
\includegraphics[width=\linewidth]{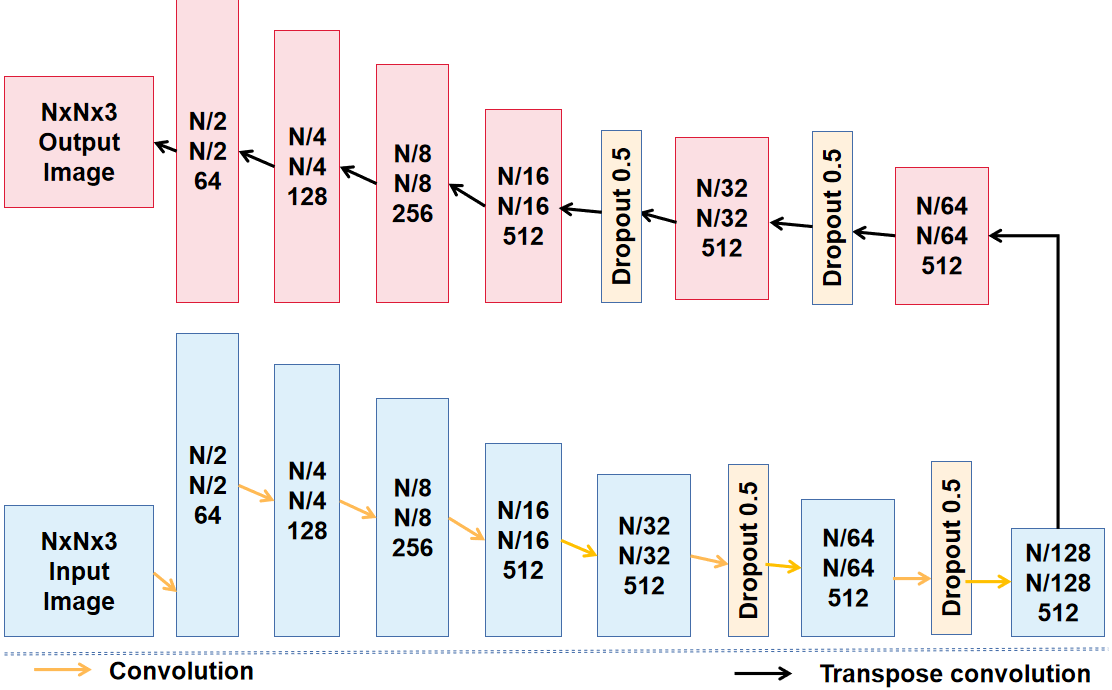}
\caption{Network details of the generator}
\label{fig:4}
\end{figure}

\begin{figure}[h]
\centering
\includegraphics[scale = 0.40]{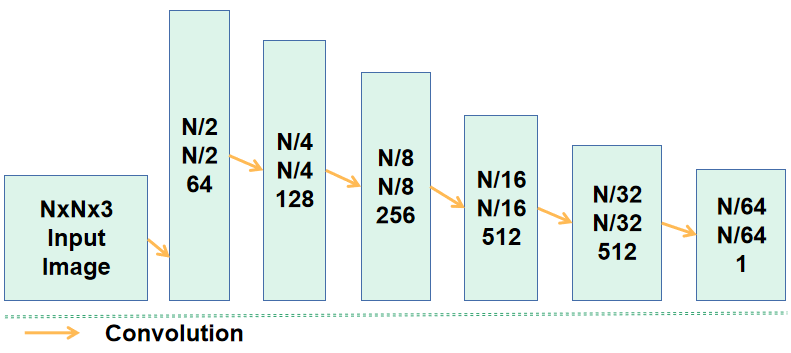}
\caption{Network details of the discriminator}
\label{fig:5}
\end{figure}

Patel et al. \cite{patel2017generative} came up with a novel GAN to learn how to map a real-world luminance to the luminance of the output equipments. The author built their dataset from 20 data sources, but some of them do not have labels. In a supervised manner, it is necessary to get the true labels. To solve this problem, they carried out 12 TMOs to obtain 12 tone-mapped WDR images for each. Only one image is needed as a label, so they picked up one tone-mapped image with the highest tone mapped image quality index (TMQI) score, a metric combines a multi-scale structure fidelity measurement with an image naturalness measurement and provides a single quality score for the entire image, and discarded other 11 images. In this method, the networks were created to learn from existing tone mapping operators that are assigned to generate the GT, to create an optimized combination of them. They explored three different networks including a CNN model, a GAN, and a GAN with skip-connections. The CNN is the most basic and naive neural network (NN) as shown in Fig. \ref{fig:4}, containing downsampling and upsampling with a mean square error as the loss function. The generator of the second NN has the same architecture as CNN in the first method. It executed an L1 loss between the real and generated images. Between the generator and discriminator, a binary cross-entropy loss function was harnessed. This GAN produced results that were superior to that of the CNN, because the discriminator is capable of learning the distribution of the GT that guides the generator producing more vivid images. The discriminator utilized a conventional CNN as shown in Fig. \ref{fig:5}, limited to six convolutional layers for downsampling, in order to retain freedom to the discriminator in differentiating true label or generated images. The third network introduced skip connections between corresponding upsample and downsample layers to solve the lossy in output clarity that happened in the GAN network. The addition of skip connections maintained signal strength to avoid information loss at pixel level from downsampling layers, resulting in a higher quality output than the second network. This operator learns from many prevalent TMOs to make visually pleasing images, just because of this, its performance would not exceed these TMOs significantly.\\

\begin{figure}[t]
\centering
\includegraphics[width=\linewidth]{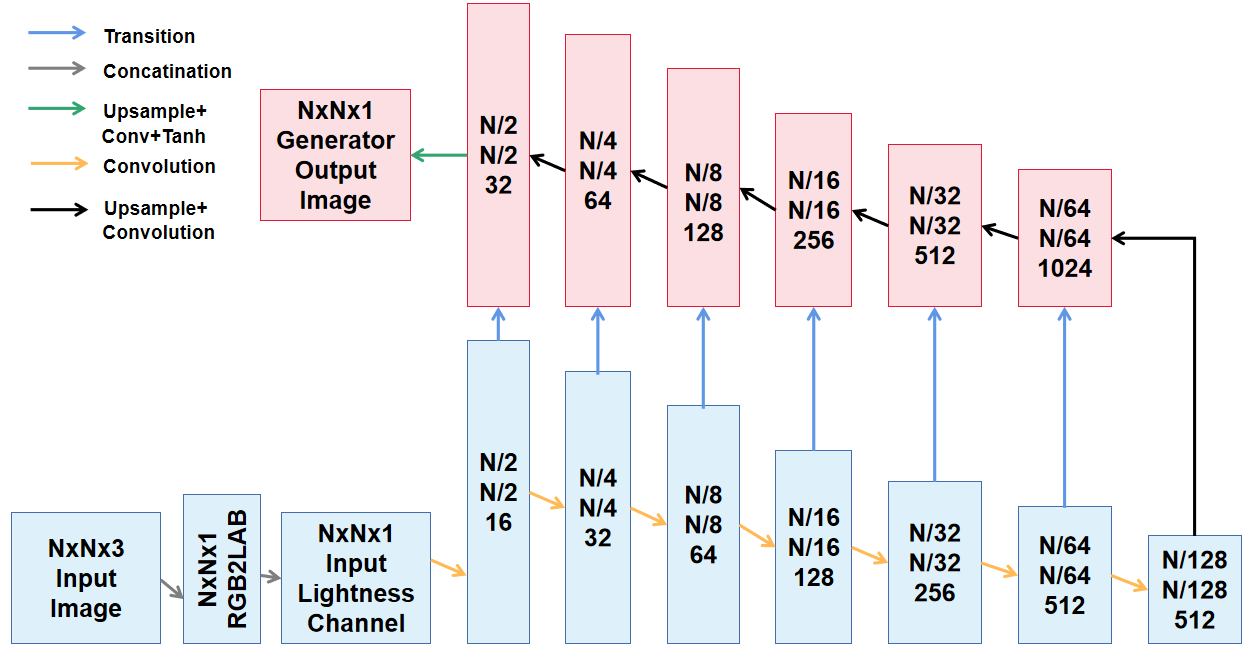}
\caption{DCGAN U-net Generator network architecture}
\label{fig:6}
\end{figure}

\begin{figure}[b]
\centering
\includegraphics[scale = 0.30]{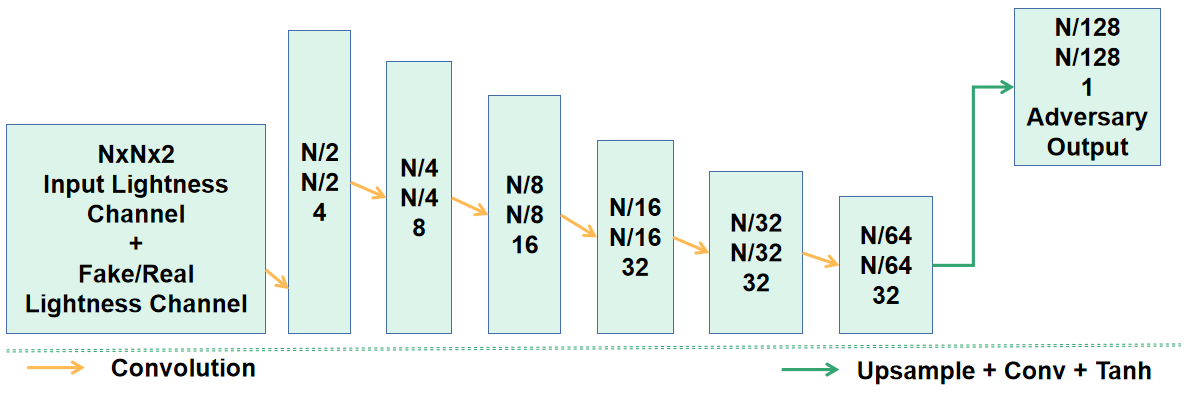}
\caption{DCGAN Discriminator network architecture}
\label{fig:7}
\end{figure}

\begin{figure}[!h]
\centering
\includegraphics[width=\linewidth]{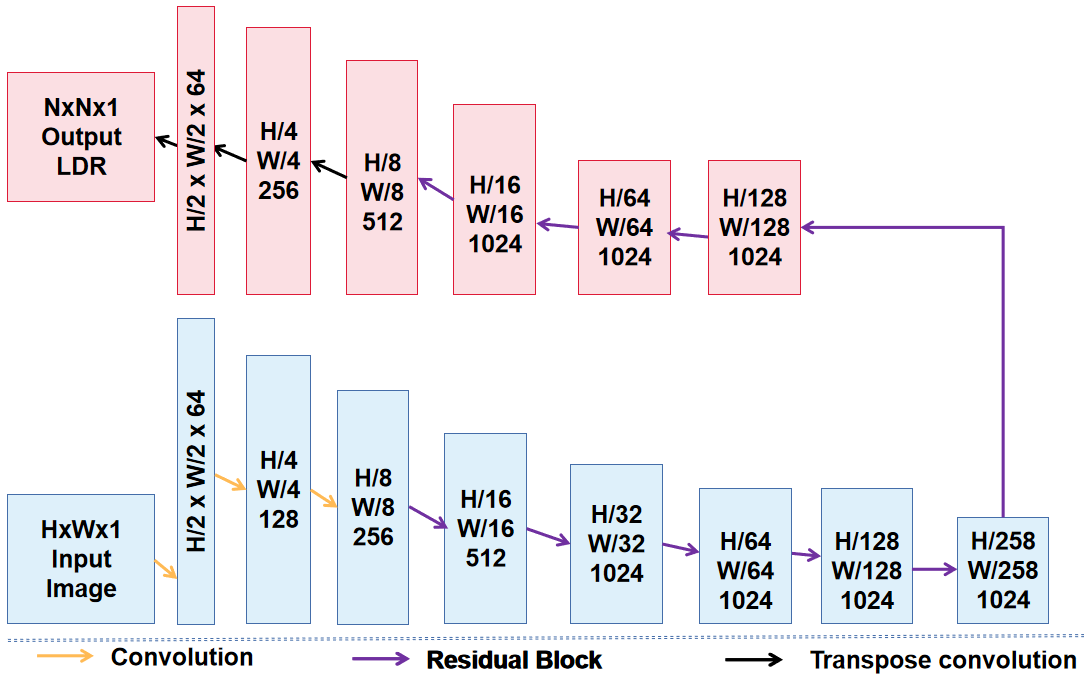}
\caption{Generator (Single Scale)}
\label{fig:8}
\end{figure}

\begin{figure}[!h]
\centering
\includegraphics[scale = 0.30]{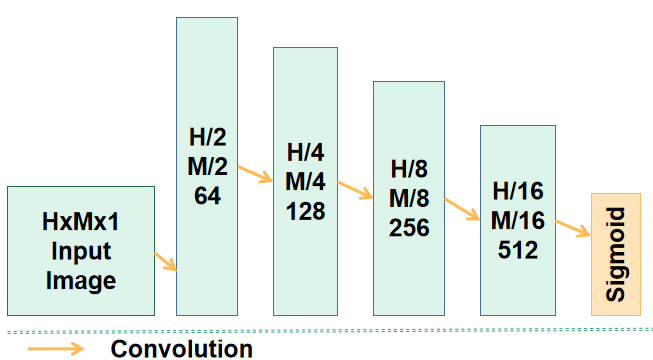}
\caption{Discriminator (Single Scale)}
\label{fig:9}
\end{figure}

\begin{figure}[!h]
\centering
\includegraphics[scale = 0.30]{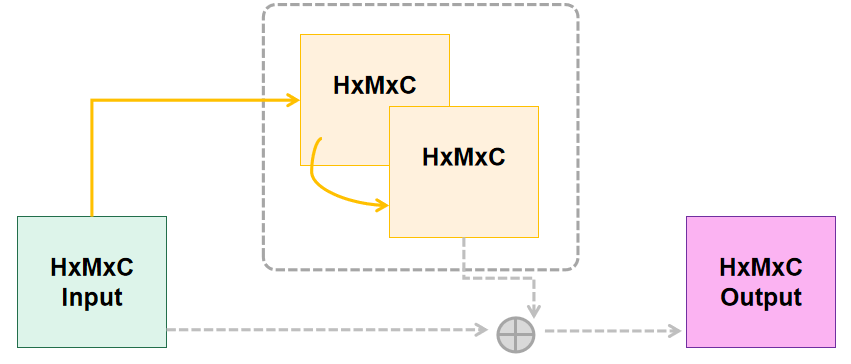}
\caption{Residual block}
\label{fig:10}
\end{figure}

Montulet et al. \cite{montulet2019deep} focused on the utilization of GAN for low-light image enhancement, applying deep convolutional generative adversarial network (DCGAN) to the luminance channel instead of directly to color components, in order to avoid color bias. A U-net is exploited in the generator as shown in Fig. \ref{fig:6}, a Patch-GAN is adopted in the discriminator, and VGG based perception-related loss function is used in the generator. Ledig et al.  \cite{ledig2017photo} introduced VGG as their loss function on the task of super-resolution, and achieved success in practice. VGG loss also called perceptual loss is an alternative to a pixel-wise loss like MSE, which using a pre-trained model applies convolution layers to extract features and patterns from the displayed images obtained from the generator and GT labels, and reduce the gap between the outputs and GT labels. Unlike the traditional GAN model that carry outs a deep CNN as a discriminator to classify the full image, this model opts a PatchGAN as shown in Fig. \ref{fig:7}, which is designed to classify whether a part of the input image is real or fake. The number of layers in PatchGAN is configured so that the effective receptive field of each output is mapped to a specific size in the input image. The output of the network is a single feature map with real/fake prediction values. The values of this map can be averaged to give a score. Montulet et al.  \cite{montulet2019deep}’s method is decided on the traditional TM methods to generate the target LDR images, so their behavior will not be noticeably different from that of the traditional TM method.\\
Rana et al. \cite{rana2019deep} shows a similar advantage of GAN over CNNs, since convolutional 
neural networks can result in widely varying output qualities depending on the choice 
of a loss function. Due to the dependence of tone mapping operators on well-adjusted 
parameters, the authors created DeepTMO, a parameter-free operator based on CGAN. An advantage of using a CGAN for tone mapping compared to a GAN is that the CGAN is constrained by each output pixel’s conditional dependence on at least one neighboring pixel in the input. This results in the network receiving a penalty for structural difference between input and output, which is a desirable quality in tone mapping, where the output should remain effectively visually similar to the input. The drawback of using a CGAN is that it is more challenging to generate a high-resolution resulting image due to instability and problems with optimization. The method involves four CGAN settings which explored possible combinations of single-scale and multi-scale generators and discriminators. Fig. \ref{fig:8} displays a single-generator harnessing an encoder-decoder architecture, while the single-discriminator is essentially a PatchGAN shown in Fig.      \ref{fig:9}, chosen for its simplicity over a full-size discriminator, with fewer parameters. A PatchGAN discriminator can be used for spatially localized improvement as well. Fig. \ref{fig:10} presents a residual block that consists of two sequential convolution layers.  \\

\begin{figure}[h]
\centering
\includegraphics[width=\linewidth]{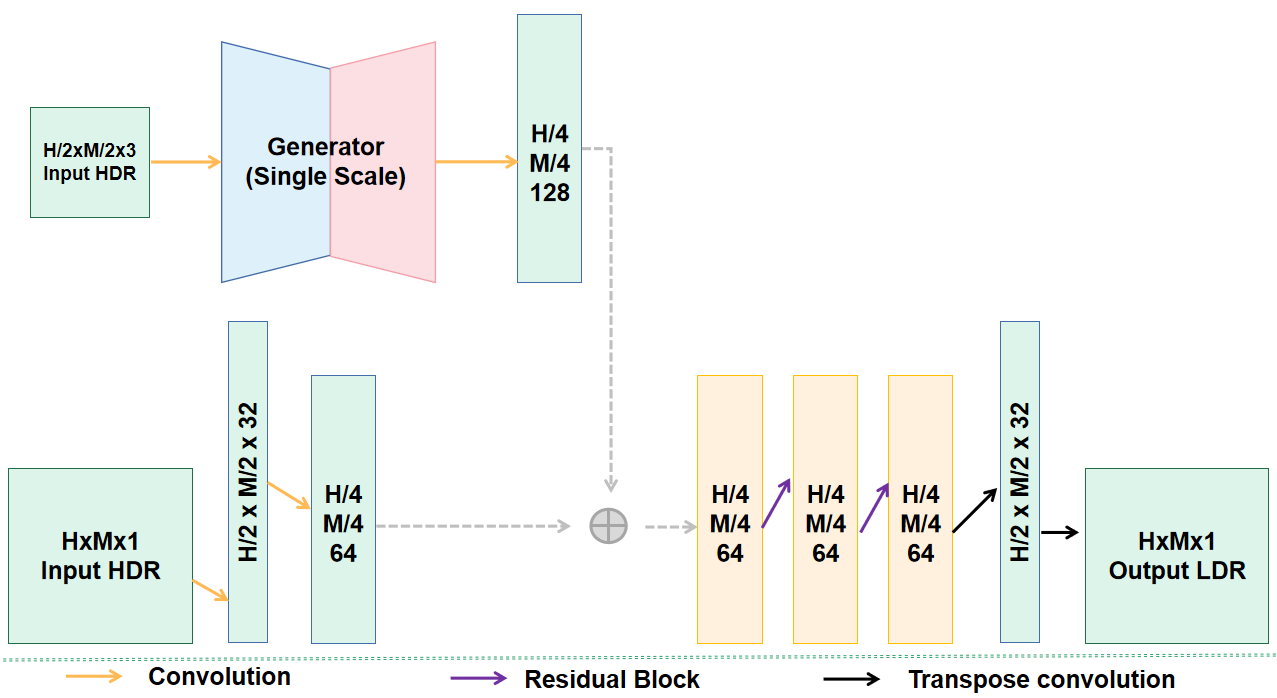}
\caption{DeepTMO multi-scale generator architecture}
\label{fig:11}
\end{figure}

A single-generator, single-discriminator gives acceptable global-level reconstruction, but the output has high levels of noise in some specific sections of results. A multi-generator with a multi-discriminator network gives an output of strong quality with no artifacts. A multi-discriminator is fundamentally the same as the single-discriminator, but harness it on two scales of input. For a multi-generator shown in Fig. \ref{fig:11}, it consists of a global original network and a down-sampled network. The down-sampled model is similar to the single-generator, and its generated images are inputted into the global network to combine coarse and fine-grained details together. The combinations of single-generator with multi-discriminator and multi-generator with single-discriminator, give results landing in between the full single or completely muti-structures. The dynamic range is initially compressed in the luminance channel, and colors are reproduced in post-processing operations. This TMO could cater to vast scenic-content (e.g., outdoor, indoor, landscapes, structures, human, etc.), and solve the problems (e.g., saturation, pattern blurring, tiling patterns) caused by traditional GAN-based TMOs \cite{montulet2019deep}, but when input images have a different size from the training image, it could perhaps result in artifacts.\\

\begin{figure}[h]
\centering
\includegraphics[width=\linewidth]{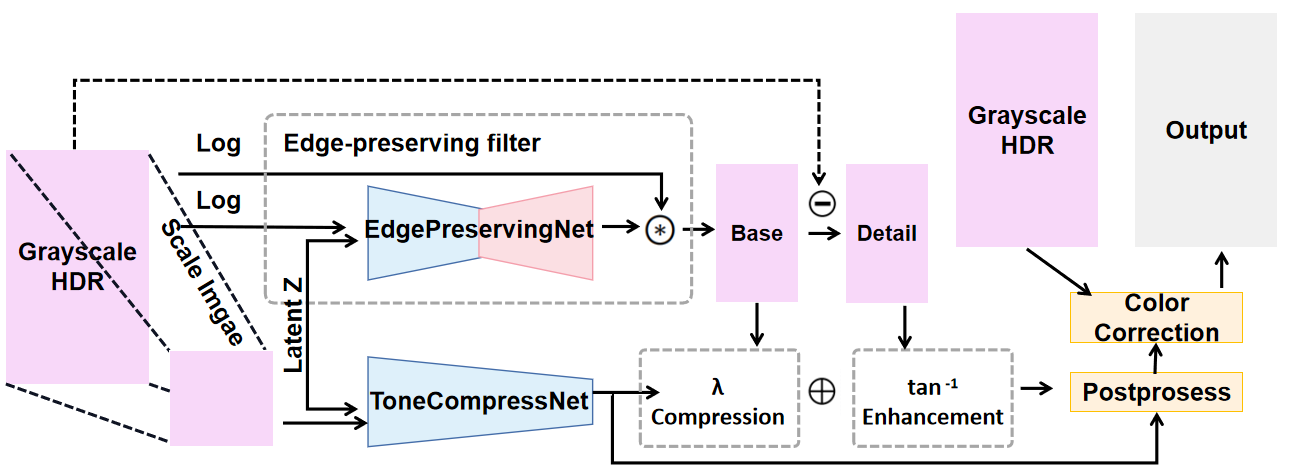}
\caption{Overall block diagram of Su et al.'s \cite{su2020explorable} method}
\label{fig:12}
\end{figure}

\begin{figure}[h]
\centering
\includegraphics[width=\linewidth]{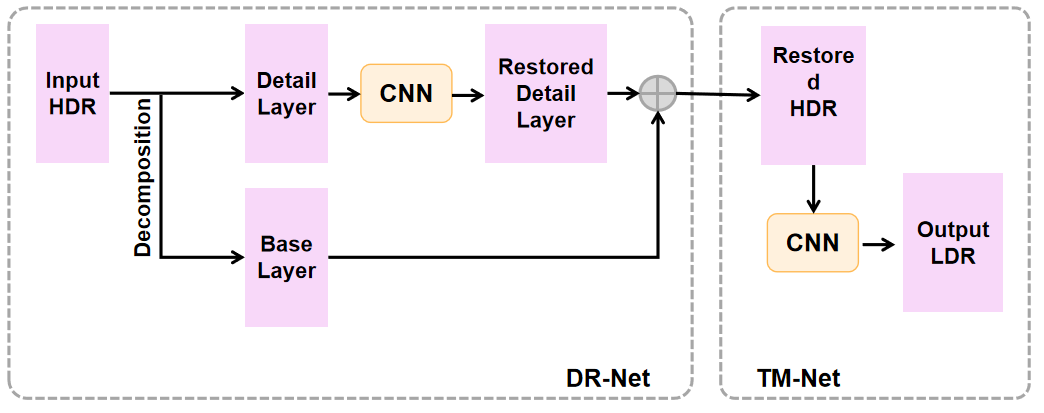}
\caption{Kim et al.'s \cite{kim2020detail} tone-mapping method}
\label{fig:13}
\end{figure}

Su et al. \cite{su2020explorable} developed a multimodal tone-mapping method composed of two networks called EdgePreservingNet and ToneCompressingNet in Fig. \ref{fig:12}. EdgePreservingNet is trained to filter the input WDR images, preserving the high-frequency information in the image. The WDR images are separated into a detail and a base layer. The detail layer is enhanced as a result of a tan function for producing better quality LDR images, meanwhile the base layer is squeezed via ToneCompressingNet. Ultimately, the enhanced detail layer and compressed base layer are fused into the final grayscale LDR output. Through the color correction, generating the final colorful tone-mapped WDR image. The author determined U-Net as the architecture of  EdgePreservingNet. For ToneCompressingNet, it contains a serial continuous convolution layers and a fully connection layer in the end. \\
Kim et al. \cite{kim2020detail} suggested a TMO for X-ray images, which could be divided into two parts (see in Fig. \ref{fig:13}). Detail-recovery network (DR-Net) is responsible for restore details in the WDR image, while TM-Net (TM-Net) aims for compressing dynamic range. This is the first learning-based TMO used for X-ray inspection. For training DR-Net, the author presented a data synthesis technique, which is based on the Beer-Lambert law \cite{barrett1996radiological}, to generate ground-truth (GT). DR-Net utilized a guided filter \cite{he2012guided} to decompose the input WDR images into 2 layers: a detail layer and a base layer. The detail layer is chosen for restoring missing details, then this layer is integrated with the base layer. After restoring details by DR-Net, the output LDR image is passed to TM-Net. However, there does not exist standard X-ray databases for TM, leading to that it is impossible to train TM-Net in a supervised training method. To solve the problem, Kim presents a novel loss function named structural similarity loss, which not only enhances details, but also alleviates the halo artifacts. Therefore, the author chooses U-Net as the basis of DR-Net and TM-Net.\\

\section{Conclusion}
Evidently, the usage of GANs for tone mapping produces desirable results. GANs overcome the limitations of CNNs. As the popularity of GAN rapidly rises and the necessity for well-mapped images increases, perhaps their intersection will find an optimized display of every type of image.
Notwithstanding TMQI metric is a good measurement to compare outputs to existing operators, it can also restrict the network, as it is only an approximation to the HVS. To address this problem, having a stronger formulation of what makes an image look the “best” to most people, beyond noise and artifacts, would a strong start. Nowadays, traditional TM algorithms are surpassed by machine learning models. What is more, look at the new displays that have more than 8 bits in their display, (e.g. the new Apple device), and conclude that in the near future we believe tone mapping will not be required anymore due to the display devices.

\ifCLASSOPTIONcaptionsoff
  \newpage
\fi



%

\bibliographystyle{IEEEtran}
\bibliography{references}

\end{document}